\documentclass[prd,twocolumn,groupedaddress,showpacs,nofootinbib]{revtex4}
\pdfoutput=1
\usepackage{graphicx}
\usepackage{bm}
\usepackage{times}
\usepackage{slashed}
\usepackage{color}
\usepackage{color}

\begin{document}

\title{Stringent Dilepton Bounds on Left-Right Models using LHC data}
\author{Sudhanwa Patra$^{1,2}$, Farinaldo S.\ Queiroz$^{1}$, Werner Rodejohann$^{1}$}
% \email{sudha.astro@gmail.com}
\email{sudhanwa@mpi-hd.mpg.de}
\email{queiroz@mpi-hd.mpg.de}
\email{werner.rodejohann@mpi-hd.mpg.de}
\affiliation{$^{1}$Max-Planck-Institut f\"ur Kernphysik, Postfach 103980, 69029 Heidelberg, Germany \\
$^{2}$ Center of Excellence in Theoretical and Mathematical Sciences, \\
%\hspace*{-0.2cm} 
Siksha 'O' Anusandhan University, Bhubaneswar-751030, India\\
%$^{3}$ Department of Physics and Astronomy, Mitchell Institute for Fundamental Physics and Astronomy, Texas A \& M University, College Station, TX 77843-4242.\\
%$^{4}$Department of Physics and Santa Cruz Institute for Particle Physics University of California, Santa Cruz, CA 95064, USA
}
\begin{abstract}
\noindent 
In canonical left-right symmetric models the lower mass bounds on the charged gauge bosons are in the ballpark of $3-4$ TeV, resulting into much stronger limits on the neutral 
gauge boson $Z_R$, making its production unreachable at the LHC. 
However, if one evokes different patterns of left-right symmetry breaking the 
$Z_R$ might be lighter than the $W_R^\pm$ motivating an independent $Z_R$ collider study. In this work, we use the 8 TeV ATLAS $20.3$ fb$^{-1}$ luminosity data 
to derive robust bounds on the $Z_R$ mass using dilepton data.  
%because they provide the most restrictive limits due to the sizable $Z_R$-lepton couplings. 
We find strong lower bounds on the $Z_R$ mass for different right-handed gauge couplings,   excluding $Z_R$ masses up to $\sim 3.2$~TeV. For the canonical LR model we place a lower mass bound of $\sim 2.5$~TeV. Our findings are almost independent of the right-handed neutrino masses ($\sim 2\,\%$ effect) and applicable to general left-right models.
\end{abstract}

\maketitle
%%%%%%%%%%%%%%%%%%%%%%%%%%%%%%%

\noindent
\section{Introduction} 
Left-Right (LR) symmetric models are popular extensions of the 
Standard Model (SM) and are based on the gauge group 
$SU(2)_L\otimes SU(2)_R\otimes U(1)_{B-L}$ \cite{LR}. They were 
originally motivated to explain the origin of parity violation \cite{Senjanovic} of weak interactions 
and found to be related to the generation of light neutrino masses via the 
seesaw mechanisms \cite{typeI,typeII}, linking in fact the smallness of neutrino mass with parity 
violation. The LR symmetry may be interpreted as the first step of an eventual unification of 
gauge forces as well.

While aesthetically very appealing, the theories do not predict the scale of parity 
restoration, leaving this question open to experiment. What is generic in LR symmetric models is the 
presence of right-handed currents and of gauge bosons $W_R^\pm$ and $Z_R$ 
associated with the additional $SU(2)$ gauge group. 
The search for those generic features has been however unsuccessful so far. 
Recently, several studies have been made in TeV scale LR symmetric models exploiting meson \cite{Beall:1981ze,Bertolini:2014sua}, collider \cite{Bakhet:2014ypa,Chen:2013foz,Nemevsek:2012iq,CMS:2012zv}, flavor \cite{Castillo-Felisola:2015bha,Dekens:2014ina,Das:2012ii} and neutrinoless double beta decay data 
\cite{Barry:2013xxa,Huang:2013kma,Awasthi:2013ff,Chakrabortty:2012mh,Dev:2013vxa}

What these analyses have in common is assuming that (i) 
the masses of the charged bosons $W_R^\pm$ are smaller than that of 
the neutral one $Z_R$, and that (ii) the gauge couplings of the left- and right-handed interactions 
($g_L$ and $g_R$) are identical. This implies in particular that the effects of the $W_R^\pm$ are 
the ones that matter in testing the models and in determining the scale of LR symmetry. 
Indeed, models typically advocate the presence of triplet and bidoublet scalars to generate the 
fermion and gauge boson masses. In this case both $g_L=g_R$ and $M_{W_R} > M_{Z_R}$ result, 
and the scenario has been widely explored with TeV scale breaking of 
$SU(2)_{R} \times U(1)_{B-L}$ down to $U(1)_Y$ resulting 
in  charged and neutral gauge bosons with masses around the TeV scale.  
Within this symmetry breaking pattern $W_R$ 
collider searches have been applied because they provide stronger constraints. 
For instance, CMS imposes $M_{W_R} \gtrsim 3$ TeV \cite{Khachatryan:2014dka} 
assuming $M_{W_R} > M_{N_R}$ ($N_R$ being the right-handed neutrinos), 
which translates into $M_{Z^\prime} \gtrsim 5.1$~TeV using the mass relation 
$M_{Z_R} \simeq 1.7 M_{W_R}$ that holds in canonical LR models. There are also important limits stemming from electroweak 
data ($M_{Z_R} \gtrsim 1$~TeV), and $K-\bar{K}$ oscillations ($M_{W_R} \gtrsim 4$~TeV for P-parity as 
the discrete LR symmetry), which results into $M_{Z_R} \gtrsim 6.8$ TeV \cite{Bertolini:2014sua}. 

In this respect, one can compare the limits with $Z^\prime$ constraints of theories in which only 
$B-L$ is gauged. LEP2 imposes $M_{Z^\prime} \gtrsim 6 \times g_{BL}$~TeV, where 
$g_{BL}$ is the gauge coupling \cite{Carena:2004xs}. 
However, this limit on the $B-L$ coupling cannot be easily applied to 
LR models, even though there is a gauge coupling relation $1/g_Y^2 = 1/g_{BL}^2 + 1/g_R^2$. 
In $U(1)_{B-L}$ theories $g_{BL}$ solely determines the $Z^{\prime}$-fermion couplings which are purely 
vectorial. In LR models, on the other hand, there are vector and axial couplings and other 
constant factors related to the Weinberg angle which suppress the coupling to leptons. 
Therefore, those extra factors should be taken into account. Indeed, the overall couplings to leptons are 
dwindled along with the LEP2 bound stemming from $B-L$ theories, as one can see in Ref.
\cite{delAguila:2010mx}. There, the authors found a lower mass limit of $667$~GeV on the 
$Z^{\prime}$ for $g_R=g_L$.

We stress here the both features mentioned above, $g_L=g_R$ and $M_{W_R} > M_{Z_R}$, are not guaranteed 
in general, and the experimental searches for LR symmetry should not be limited to those assumptions. In particular one might evoke different symmetry breaking patterns yielding $M_{Z_R} < M_{W_R}$. Hence from a general perspective, it is crucial to carry out an independent $Z_R$-collider study, since in this mass regime $Z_R$ collider searches become the most effective way of constraining LR models specially when $W_R$ is sufficiently heavy, out of reach of current experiments. Without losing generality, 
we use in this paper ATLAS data at 8 TeV and $20.3\mbox{ fb}^{-1}$ luminosity, to set limits on the $Z_R$ mass of LR models using dilepton data using MadGraph5,Pythia and Delphes3, as dilepton resonance searches are the most efficient method to bound neutral gauge bosons 
that have non-negligible couplings to leptons \cite{Arcadi:2013qia,Alves:2015pea}.  We emphasize that our results are quite general because they rely only on the neutral current of the $Z_R$ gauge boson.

Additionally to those limits, we present as an explicit example a model based on a two step-breaking pattern which generates $M_{Z_R} < M_{W_R}$ in a consistent way, with predicted $g_R/g_L$ ratios by forcing unification at the GUT scale. 

%Our paper is structured as follows: firstly, we introduce the LR symmetric model with a  
%two-step breaking leading to $M_{Z_R} < M_{W_R}$ (a potential $SO(10)$ embedding and gauge coupling analysis 
%is delegated to the Appendix); secondly we derive the dilepton limits on $Z_R$ using 
%ATLAS 8 TeV with $20.3\mbox{ fb}^{-1}$ luminosity data; 
%lastly we draw our conclusions.

 %%%%%%%%%%%%%%%%%%%%%%%%%%%%%%%%%%%%%%%%%%%%%%%%%%%%%%%%%%%%%%%%

\section{Left-Right Symmetric Model\label{sec:mod}}
Left-right symmetric models 
are based on the gauge group $SU(2)_L \times SU(2)_R \times U(1)_{B-L} \times SU(3)_C$. In addition, 
a discrete left-right symmetry is present implying equal values of gauge couplings for the  $SU(2)_{L,R}$ 
gauge groups i.e.\ $g_{L}=g_{R}$. The quarks and leptons come as LR symmetric doublet 
representations $Q_{L,R}=(u,d)^T_{L,R}$ and $\ell_{L,R}=(\nu,e)^T_{L,R}$. In the conventional and most often 
studied left-right symmetric models 
$SU(2)_R \times U(1)_{B-L}$ is broken down to $U(1)_Y$ in one single step. In particular, 
the discrete left-right symmetry (denoted as parity or charge conjugation) and the 
$SU(2)_R$ gauge symmetry are broken at the same time and scale\footnote{It is possible to break the 
discrete and gauge symmetry at different scales, leading in particular to 
$g_L \neq g_R$ at the electroweak scale \cite{Chang:1983fu,Chang:1984uy}, see also the Appendix.}. 

However, it should be noted here that the spontaneous symmetry breaking of 
$SU(2)_{R} \times U(1)_{B-L}$ down to $U(1)_Y$ can be achieved 
either by Higgs triplets ($\Delta_{L} \oplus \Delta_{R}$ with even $B-L=2$) or 
Higgs doublets ($\chi_{L} \oplus \chi_{R}$ with odd $B-L=-1$) or a combination of both Higgs 
doublets and triplets. With the simple Higgs sector comprising of a bidoublet plus $SU(2)_{L,R}$ 
triplets $\Delta_{L,R}$, the known formula between the right-handed charged and neutral gauge boson masses is given by 
\begin{eqnarray}
\frac{M_{Z_{R}}}{M_{W_{R}}}=\frac{\sqrt{2} g_R/g_L}{\sqrt{(g_R/g_L)^2-\tan^2\theta_W}}\,.
\end{eqnarray}
With $g_L\simeq g_R$, one can find that $M_{Z_{R}}=1.7 M_{W_{R}}$. Thus, the existing experimental bounds on $M_{W_R}$ can be translated into more restrictive limits on $M_{Z_{R}}$. The aforementioned limits on the $W_R$ mass which are in the ballpark of several TeV make the $Z_R$ production unattainable at the LHC.

Albeit, in this work we consider different classes of LR models, where this mass relation 
does not apply. In particular, the $W_R$ mass is set to be at a scale much larger than TeV, 
whereas the $Z_R$ mass lies at the TeV scale. A possible way to conceive this setup is 
by introducing two triplet scalars $\Omega_{R,L}$ and $\Delta_{L,R}$, with $B-L=0,-2$ respectively. With their inclusion the LR symmetry breaks down to the 
SM gauge group in two steps: (i) $SU(2)_L \otimes SU(2)_R \otimes U(1)_{B-L}$ breaks 
to\footnote{Note that the rank of $SU(2)_R$ and $U(1)_R$ are the same, which always happens if 
the breaking is mediated by a Higgs field in the adjoint representation.}  
$SU(2)_L \otimes U(1)_R \otimes U(1)_{B-L}$ at $W_R$ mass scale which is implemented through 
the vacuum expectation value (vev) of the heavier triplet carrying $B-L=0$, 
i.e.\ $\Omega_R$; (ii) then $U(1)_R \otimes U(1)_{B-L}$ 
breaks down to $U(1)_Y$ at the $Z_R$ mass scale defined by the vev of 
the $\Delta^0_R$. Setting the vev of $\Omega_R$ to a very high energy scale, $W_R$ completely decouples 
from $Z_R$. Moreover, we need a bi-doublet ($\Phi$) to break $SU(2)_L \otimes U(1)_Y$ down to electromagnetism. 

Assuming the lighter right-handed neutral gauge boson $Z_R$ acquires mass at the TeV scale, then $Z_R$ searches 
become the most promising ones to derive limits on the mass of this neutral boson. 
In the next section we discuss the $Z_R$ phenomenology and derive 
dilepton limits using recent ATLAS data.

Of course, one does not have to rely on the precise symmetry breaking pattern we are proposing. 
The idea of having a light neutral gauge boson and a very heavy charged one is what our reasoning is 
mostly based on. A more detailed study of the above symmetry breaking pattern, plus analyses of 
the scalar potential, neutrino masses and other phenomenological consequences will be presented 
elsewhere \cite{ModelWerner}. As far as dilepton bounds are concerned which are the focus of this work, 
the relevant interactions for us are the $Z_R$-fermions couplings given by
\begin{eqnarray} \label{eq:coupl}
\frac{g_R}{\sqrt{1-\delta \tan^2\theta_W}}
\overline{f}\,\gamma_\mu \left(
     g_V^f - g_A^f \gamma^5 \right) \,f\,\, Z_R^{\mu} \,,
\end{eqnarray}
where the vector and axial couplings are defined 
as 
\begin{eqnarray}
&&\hspace*{-0.5cm}g^f_V=\frac{1}{2}\left[\big\{\delta \tan^2\theta_W \left(T^{f}_{3L} 
     -\mbox{Q}^f \right)\big\} 
+ \big\{T^{f}_{3R}- \delta \tan^2\theta_W \mbox{Q}^f\big\}\right] \nonumber \\
&&\hspace*{-0.5cm}g^f_A=\frac{1}{2}\left[\big\{\delta \tan^2\theta_W \left(T^{f}_{3L} 
     -\mbox{Q}^f \right)\big\} 
- \big\{T^{f}_{3R}- \delta \tan^2\theta_W \mbox{Q}^f\big\}\right] \nonumber
\end{eqnarray} 
with $\delta=g^2_L/g^2_R$ the ratio of gauge couplings, $T^{f}_{3L,3R}$ being $1/2$ ($-1/2$) 
for up (down) type fermions and $Q^f$ the respective electric charge. It is clear that the neutral current is general, and thus the coupling strengths. For this precise reason the limits that we are about to derive in the next section are applicable to a rather large class of LR models.

\begin{figure}[!t]
\includegraphics[scale=0.5]{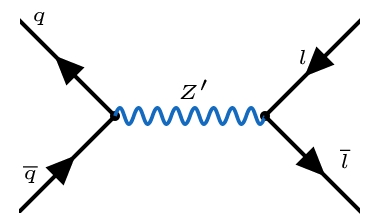}
\caption{Feynman diagram representing dilepton resonance searches at LHC.}
\end{figure}

\section{Dilepton Limits} 

\begin{figure}[!t]
\includegraphics[scale=0.5]{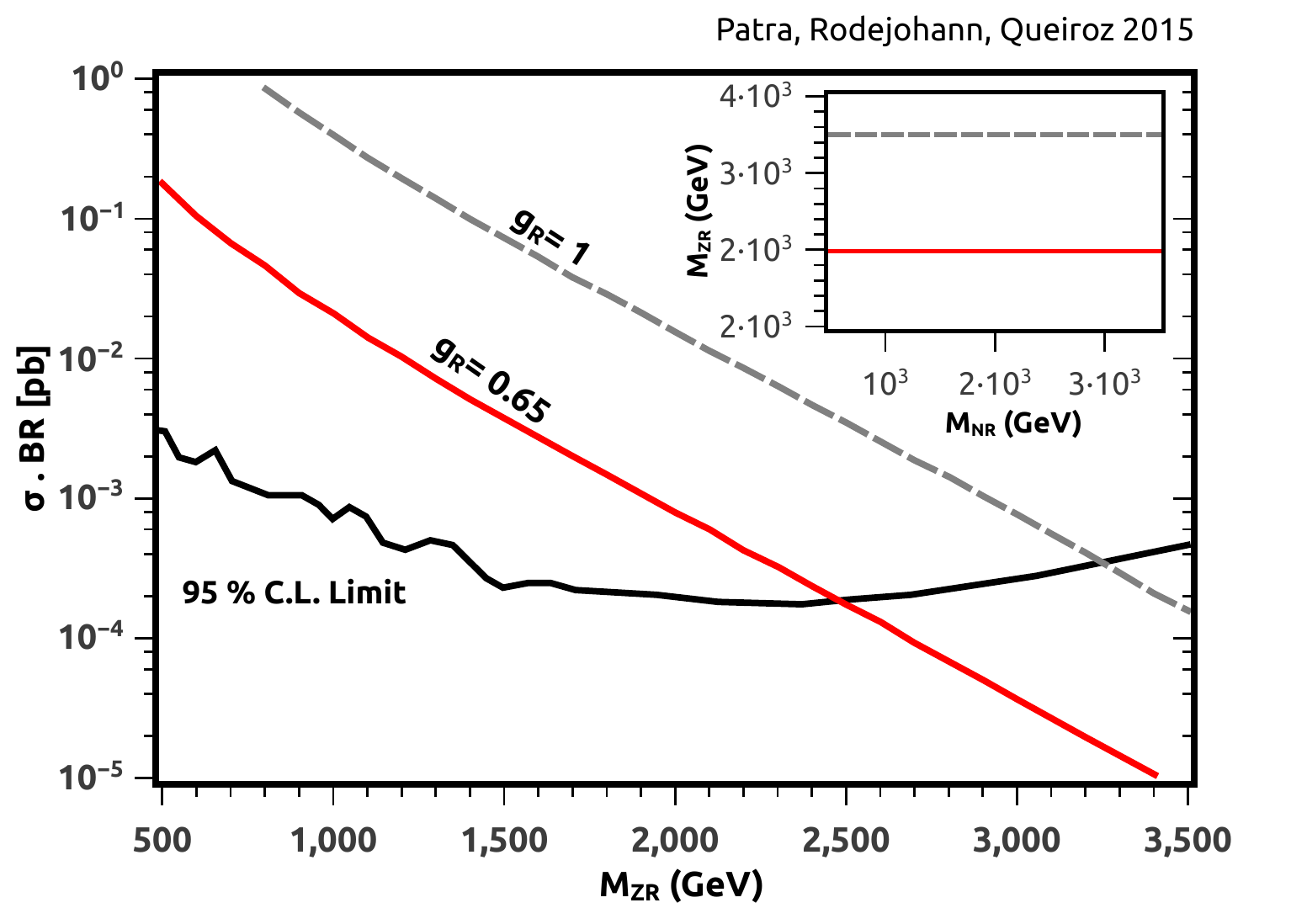}
\caption{Dilepton limits on the $Z_R$ mass using ATLAS 8 TeV 20.3 $\rm fb^{-1}$ integrated luminosity data for two different $g_R$ values ($g_R = 0.65$ is the canonical value equal to $g_L$). For a larger $g_R$ range see Fig.\ \ref{fig:2}. In the inner graph we show the essentially  absent dependence of our limits on the right-handed neutrino masses due to their small associated branching ratio.}
\label{fig:1}
\end{figure}
%\end{center}
%\end{widetext}

%\end{center}
%\end{widetext}

Dileptons and dijet searches have been proved to be the most effective as far as bounds on 
additional $Z'$ bosons are concerned unless they have negligible couplings to leptons or 
large branching ratios to missing energy such as dark matter 
\cite{Alves:2013tqa,deSimone:2014pda,Buchmueller:2014yoa,Frandsen:2012rk}. Hence, in the classes 
of LR symmetric models we are studying we need to consider those in order to set limits on the $Z_R$. 
The neutral gauge bosons can be produced in the $s$-channel from $q\bar{q}$ annihilation. To perform the search, the dilepton invariant mass line shape is studied for a localized excess of events corresponding to a new physics resonance. 

To derive the dilepton limits on this specific model we simulate the process 
$pp \rightarrow Z_R \rightarrow e^+e^-, \mu^+ \mu^-$, plus up to two extra jets
using MadGraph5 \cite{Alwall:2011uj}, and compare with the results from the ATLAS collaboration reported in Ref.\ \cite{Aad:2014cka}, from where we also take the background events. For this reason we obtain the number of events in bins of the dilepton invariant mass $M_{ll}$ as follows: $110-200$~GeV, $200-400$~GeV, $400-800$~GeV, $800-1200$~GeV, $1200-3000$~GeV, $3000-4500$~GeV. For the signal events we account for clustering and hadronizing jets as well as for soft and collinear
QCD radiation with Pythia \cite{Sjostrand:2006za}, and simulate detector efficiencies with Delphes3 \cite{deFavereau:2013fsa}. In our results we used the CTEQ6L parton distribution functions computed at  $\mu_F = \mu_R = M_{Z_R}$ \cite{Lai:2009ne}. Following the procedure in  Ref.\ \cite{Aad:2014cka}, the signal events were selected by applying the cuts:

\begin{figure}[!t]
\includegraphics[scale=0.5]{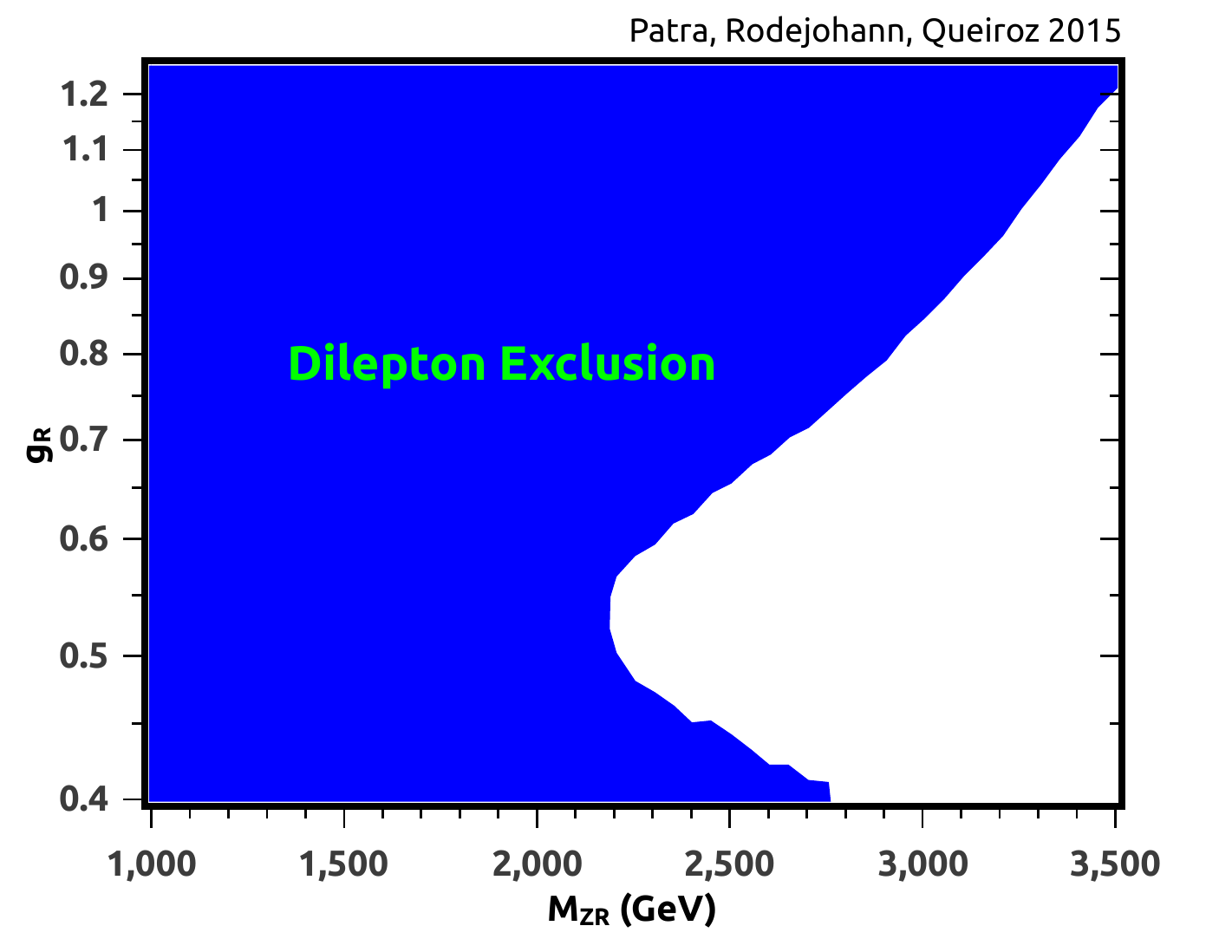}
\caption{Dilepton limits on the $Z_R$ mass for different $g_R$ values. From Eq.\ (\ref{eq:coupl})  the $Z_R$-fermion coupling strength does not always grow with $g_R$ because of the presence of extra $1/g_R^2$ factors in the vector/axial couplings, explaining the shape of the figure. %Hence, we conclude that in the canonical LR model $g_R=g_L=0.65$ we impose $M_{Z_R} > 2490$~GeV, and for $g_R= 1$ we find $M_{Z_R} > 3250$~GeV, consisting the strongest limits on the $Z_R$ mass in the literature.
}
\label{fig:2}
\end{figure}
 
\begin{itemize}
\item  $p_T(e_1) > 40 \,{\rm GeV}, p_T(e_2) > {\rm 30 \,GeV}, |\eta_e| < 2.47$,
\item  $p_T(\mu_1) > 25 \,{\rm GeV}, p_T(\mu_2) > 25 \,{\rm GeV}, |\eta_{\mu}| < 2.47$,
\item $1	10 \,{\rm GeV} < M_{ll} < 4.5 \,{\rm TeV}$,
\end{itemize}
where $l_1$ and $l_2$ represent the hardest and next hardest lepton in the event, whereas $M_{ll}$ is the 
invariant mass of the lepton pair. That being said, we simply compute the number of dilepton events 
for the signal and compare with the background events to derive $95$\% C.L.\ limits on the $Z_R$ mass. 

The result is shown in Fig.\ \ref{fig:1} for $g_R= 0.65,1$, where the former value corresponds to the canonical LR model. For $g_R=0.65 \, (1)$ we exclude $Z_R$ masses below $2490\, (3250)$ ~GeV. We point out that our results are independent of the right-handed neutrino masses, as shown in the inner plot in Fig.\ \ref{fig:1}, simply because the branching ratio into right-handed neutrinos is rather small. When right-handed neutrinos are kinematically available for the $Z_R$ to decay into the limits on the $Z_R$ change by approximately $2\%$, which is basically unnoticeable in the inner graph of Fig.\ \ref{fig:1}. 
%We picked those values for $g_R$ simply to cover the canonical LR model and possible larger couplings regime.

In order to account for several possible symmetry breaking schemes which may induce different $g_R$ values, we show in Fig.\ \ref{fig:2} how our limits change as we vary $g_R$. We stress that as $g_R$ increases the $Z_R$-fermion couplings do not necessarily grow as one can see from Eq.\ (\ref{eq:coupl}), since there are additional $1/g_R^2$ factors in the vector/axial couplings, explaining the shape of the figure, which is different from the one with $W_R$ bounds discussed in Ref.\ \cite{Nemevsek:2011hz}. From Fig.\ \ref{fig:2} we observe that dilepton data excludes $Z_R$ masses below $2760,2209,2314,2643$~GeV for $g_R=0.4,0.5,0.6,0.7$ respectively. 
%We stopped at $g_R$ close to unit in Fig.\ \ref{fig:2} because of the oddness of having such large gauge couplings. 
So far those $g_R$ values are simply random choices, but we stress that by embedding the  symmetry breaking scheme in a $SO(10)$ model, $g_R$ can be predicted by enforcing gauge coupling unification as shown in the appendix for a particular example. 
%In particular, $g_R$ values close to 0.4 are naturally found.

In summary, our limits are quite general because they rely simply on the $Z_R$-fermions couplings and thus are applicable to a multitude of LR models. Besides, they comprise the most stringent direct limits on the $Z_R$ mass. We point out that the scale of the Left-Right symmetry breaking can always be pushed up to higher scales, in principle, to evade our bounds, i.e. assume heavier mediators. Concerning, collider projections, in order to determine the discovery potential at LHC 13 TeV for instance, one would have to know the fake jet rate and the dilepton efficiencies at LHC 13 TeV, which are unknown at this point. However, a rather speculative study could be done though. Since our limits lie in the ballpark of $2.5-3$~TeV and heavy dilepton resonances are clean signals, it is clear that $Z_R$ bosons with such masses will be either ruled out or observed at LHC 13 TeV.

%%%%%%%%%%%%%%%%%%%%%%%%%%%%%%%%%%%%%%%%%%%%%%%%%%%%%%%%%%%%%%%%%%%%%%%%%
%       LRDParity with Sequential breaking
\section{Conclusions}

Canonical left-right symmetric models suffer from strong bounds on the charged gauge boson mass, which result in much stronger limits on the $Z_R$ mass due to the mass relation that holds in those models. If one explores different patterns of left-right symmetry breaking the $Z_R$ may be light enough to be produced at the LHC while the $W_R$ is way heavier, motivating an independent $Z_R$ collider study. 

As proof of principle, we presented a symmetry breaking scheme which consistently generates the inverted mass hierarchy $M_{Z_R} > M_{W_R}$ with the $Z_R$ mass at the TeV scale. In the  appendix, we show that through demanding gauge coupling unification and embedding the model in $SO(10)$ the value of the right-handed gauge coupling $g_R$ can be predicted, which 
for the example under study is in the ballpark of $0.4$ for several $U(1)_{B-L}$ breaking  scales. We note that models with very large $W_R$ masses have the advantage of suppressing the $W_L$-$W_R$ mixing, which generates dangerously large lepton flavor violating processes. 

After showing that light $Z_R$ can be generated in LR models, we performed a collider analysis using the 8 TeV ATLAS $20.3$ fb$^{-1}$ luminosity dilepton data to derive robust and stringent bounds on the $Z_R$ mass, due to the sizable $Z_R$-lepton couplings. For different $g_R/g_L$ ratios ranging from $0.4$ up to $1.2$ to effectively cover several different patterns of symmetry breaking, our limits in the $Z_R$ mass are given in Fig.\ \ref{fig:2}. We emphasize that our results are general since they rely simply on the neutral current of the $Z_R$ gauge boson. In particular we exclude $Z_R$ masses up to $\simeq 3.2$~TeV for $g_R=1$. For $g_R=g_L$ (canonical LR model) we derive a lower masss bound of $\simeq 2.5$~TeV, which is the most stringent direct limit in the literature of LR models on the $Z_R$ mass.

Our findings are almost independent of the right-handed neutrino masses due to their small branching ratio, and applicable to general left-right models. We stress that our bounds are the leading ones when $M_{Z_R} > M_{W_R}$, and complementary to the existing indirect ones stemming from $W_R$ searches in the setup $M_{W_R} < M_{Z_R}$. Either way, we provide the most stringent direct limits on the $Z_R$ mass of LR models. 

\section*{Acknowledgements}
%{\bf Acknowledgements} 
This work is partially supported by the Department of Science and 
Technology, Govt.\ of India under the financial grant SB/S2/HEP-011/2013 (SP) and by 
the Max Planck Society in the project MANITOP (SP and WR). 
The authors thank Alexandre Alves,Carlos Yaguna, Pavel Fileviez Perez for fruitful discussions. 

%%%%%%%%%%%%%%%%%%%%%%%%%%%%%%%%%%%%%%%%%%%%%%%%%%%%%%%%%%%%%%%%%%%
\appendix
%\section{Appendix} 
\section{$SO(10)$ GUT embedding and determination of $\delta=g^2_L/g^2_R$}
For the sake of completeness, let us shortly discuss 
a possible $SO(10)$ GUT embedding of the model from Section 
\ref{sec:mod}, which consistently predicts the $g_L/g_R$ ratio and $M_{Z_R} \ll M_{W_R}$. More details will be presented elsewhere \cite{ModelWerner}. We discuss here one  exemplary symmetry breaking chain of $SO(10)$ GUT with the desire 
of having low $B-L$ breaking scale so that the extra neutral gauge boson $Z_R$ could 
be in the TeV range leading to interesting collider searches while decoupling the right-handed charged gauge bosons $W_R$.

The precise determination of $\delta=g^2_L/g^2_R$ (or $g_R$) depends upon the $SU(2)_R$ breaking scale and the choice of Higgs spectrum required for spontaneous symmetry breaking. 
Here we choose D-parity, 
which is broken spontaneously. Such LR models have been originally 
conceived in Refs.\ \cite{Chang:1983fu,Chang:1984uy} and recently in 
Refs.\ \cite{Deppisch:2014zta,Awasthi:2013ff,Borah:2013lva,Patra:2014goa}. We briefly point out here how the spontaneous D-Parity breaking scenario 
is different from usual LR model, and essentially decouples discrete and gauged left-right symmetries.  

%The advantage of considering left-right symmetric model with spontaneous D-parity 
%breaking mechanism is to provide natural explanation of maximal parity violation at weak i%nteractions 
%and explaining seesaw VEV relation consistently. 
The spontaneous breaking of D-parity occurs at reasonably high energy scale along with $SU(2)_R \to U(1)_R$ breaking, simultaneously 
resulting in a mass of $W_R$ at high scale. Below this scale, the RG evolution of gauge couplings for $SU(2)_L$ 
and $SU(2)_R$ evolves differently guaranteeing the mismatch between $g_R\neq g_L$ at low energy. At a later stage, $U(1)_R \times U(1)_{B-L} \to U(1)_Y$ breaking is achieved by $\Delta^0_R$ at the $M_{Z_R}$ scale. 
  
As a effect of spontaneous D-parity breaking mechanism, 
the RG evolution for both gauge couplings for $SU(2)_L$ and $SU(2)_R$ gauge group is 
different, resulting in different values for gauge couplings $g_L$ and $g_R$ from $M_U$ onwards up 
to $M_Z$ scale. In addition, the right-handed charged gauge bosons $W_R$ acquire mass around $\omega_R$ which we have chosen here to be greater than $10^{10}$ 
GeV making it inaccessible to high energy collider searches. We emphasize again that our bounds on the $Z_R$ mass are independent of this choice.

\begin{figure}[!t]
\includegraphics[scale=0.6]{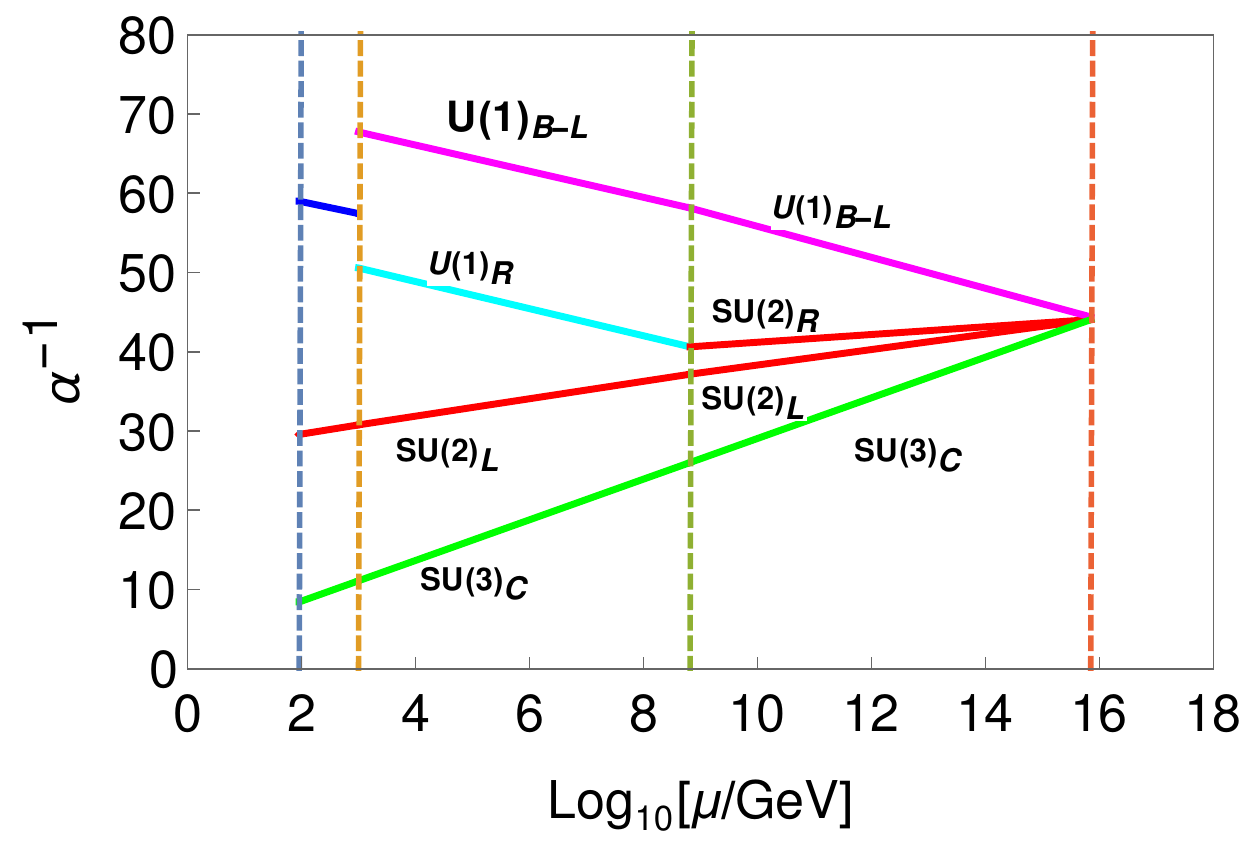}
\caption{Running of the coupling constants. By forcing grand unification at high scale we 
can predict the value of $g_R$ at the electroweak scale.}
\label{fig:unifn}
\end{figure}

We fix the $B-L$ breaking scale in the range of $1-10$ TeV to keep the $Z_R$ mass 
around LHC scale. From the unification plot for gauge couplings shown in Fig.\ \ref{fig:unifn}, 
the numerical values for the intermediate mass paramaters and the mismatch between the two gauge couplings $g_L$ and $g_R$ are estimated to be
\begin{eqnarray}\nonumber
&&M_U=10^{15.98}\,\mbox{GeV}\,,\,\, M_R=10^{9}\,\mbox{GeV}\,, \,\, M_{B-L}=5\,\mbox{TeV}\, ,\nonumber \\ \nonumber
&&\frac{g_R}{g_L} \approx 0.78\,, \quad \quad \delta=1.28\, .
\end{eqnarray}
We have checked that the coupling ratio remains basically the same for $U(1)_{B-L}$ symmetry breaking scales from 1 TeV up to 100 TeV for the present analysis. 
%Therefore, $g_R$ values around $0.4$ are natural choices. One can simply see Fig.\ref{fig:2} to find the respective dilepton bound on the $Z_R$ mass.

%%%%%%%%%%%%%%%%%%%%%%%%%%%%%%%%%%%%%%%%

\end{document}